\documentclass[nofootinbib,showpacs,preprintnumbers,onecolumn,superscriptaddress,10pt]{revtex4}
\usepackage{amsmath}

\newcommand{\vect}[1]{\mbox{\boldmath $#1$}}

\newcommand{\Lint}{{\cal L}_{\mbox{\rm\scriptsize int}}}

\newcommand{\MeV}{\mbox{\rm MeV}}
\newcommand{\GeV}{\mbox{\rm GeV}}

\newcommand{\hj}{\mbox{h\hspace{-.32em}\rule[1.25ex]
                        {.25em}{.04ex}\hspace{.07em}}}

\begin{document}


\title{Chiral symmetry breaking via constituent quarks for $\bar{q}q$
	pseudoscalar mesons}

\author{Dalibor Kekez}
\affiliation{\footnotesize Rudjer Bo\v{s}kovi\'{c} Institute,
         P.O.B. 180, 10002 Zagreb, Croatia}
 
\author{Dubravko Klabu\v{c}ar}
\affiliation{\footnotesize Physics Department, Faculty of Science,
     University of Zagreb, P.O.B. 331, 10002 Zagreb, Croatia}
 
\author{M. D. Scadron}
\affiliation{\footnotesize Physics Department, University of Arizona,
Tucson Az 85721 USA}

\begin{abstract}
We base our chiral symmetry approach on the
quark--level linear sigma model Lagrangian. Then we review
the Nambu--Goldstone theorem with vanishing $\pi$, $K$, $\eta_8$
masses. Next we dynamically generate the $\pi$, $K$, $\eta_8$ masses away
from the chiral limit. Then we study pion and kaon Goldberger--Treiman
relations. Finally we extend this $q\bar{q}$ scheme to scalar
and vector mesons. We also show the above $q\bar{q}$ meson scheme fits
the higher mass octet baryon $qqq$ pattern as well.
\end{abstract}

\maketitle

\section{Introduction}
\label{sec:intro}

Chiral symmetry breaking for pseudoscalar mesons($\pi$, $K$, $\eta_8$) requires
that although these masses vanish in the chiral limit [then satisfying
Goldberger--Treiman relations (GTRs)], $m_\pi$, $m_K$, $m_{\eta_8}$ are
non-vanishing away from the chiral limit -- hopefully near their observed
values. In a quark--level linear sigma model (L$\sigma$M) $q\bar{q}$
scheme (with constituent quarks) of Sec.~\ref{sec:QL_LsM},
the massless Nambu--Goldstone (NG) limits are reviewed in
Sec.~\ref{sec:NGtheorem} and the chiral--broken $\pi$, $K$, $\eta_8$
$q\bar{q}$ meson masses are extracted in
Secs.~\ref{sec:PiAndKmasses} and \ref{sec:Eta8}.
Then GTRs are studied in Sec.~\ref{sec:chiralGTRs}.
The analogue ground--state vector and scalar masses are obtained in
Sec.~\ref{sec:ScalarAndVector}.
Finally this L$\sigma$M $q\bar{q}$ scheme
is summarized in Sec.~\ref{sec:conclusion}.


\section{Quark--level linear $\sigma$ model}
\label{sec:QL_LsM}

   The strong interaction quark--level L$\sigma$M Lagrangian density is

\begin{equation}
\Lint
=g\bar\psi(\sigma+i\gamma_5\vect{\tau}\cdot\vect{\pi}) \psi
+g^\prime \sigma (\sigma^2+\vect{\pi}^2)
-\frac{\lambda}{4}(\sigma^2+\vect{\pi}^2)^2
-m_q \bar\psi\psi~,
\end{equation}

\noindent where

\begin{equation}
m_q=f_\pi\,g~,\,\,\,\,\,\, g^\prime=\frac{m_\sigma^2}{2f\pi}=\lambda f_\pi
\label{Consts}
\end{equation}

\noindent in the chiral limit (CL) for $f_\pi$ being approximately
$93~\MeV$ with $g=2\pi/\sqrt{3}$ \cite{Delbourgo:1993dk}.
See, {\em e.g.}, Ref.~\cite{Gell-Mann:1960np} for the original
nucleon--level version. The latter L$\sigma$M also manifests
a) the chiral current algebra
b) the fermion and meson Goldberger--Treiman relations
in Eq.~(\ref{Consts}),
c) the partially conserved axial current (PCAC) scheme as well.

\section{Review of Nambu--Goldstone theorem in CL}
\label{sec:NGtheorem}

\noindent Tree level $\partial\cdot A^{\mbox{\rm\scriptsize CL}}=0 \Rightarrow
m_\pi^{\mbox{\rm\scriptsize CL}}=0$ \cite{Bramon:1997gg}
due to the chiral symmetry extended via quark and pion loops
as Quark loops (ql):

\begin{equation}
(m_\pi^2)_{\mbox{\rm\scriptsize ql}}
=i8\,N_c\,g \left(-g + \frac{2g^\prime m_q}{m_\sigma^2}\right)
\int\frac{\hat{d}^4p}{p^2-m_q^2}
\to 0
\,\,\,\,\,
\mbox{\rm in CL}~,
\label{QuarkLoops}
\end{equation}

\noindent where $\hat{d}^4p=\frac{d^4p}{(2\pi)^4}$. Meson loops (ml):

\begin{equation}
(m_\pi^2)_{\mbox{\rm\scriptsize ml}}
=
(-2\lambda+5\lambda-3\lambda)\,
i \int\frac{\hat{d}^4p}{p^2-m_\pi^2}
+
(2\lambda+\lambda-3\lambda)
i \int\frac{\hat{d}^4p}{p^2-m_\sigma^2}
=0+0=0~.
\label{PionLoops}
\end{equation}

\noindent Then the Nambu--Goldstone theorem in the CL is

\begin{equation}
m_\pi^2
=
(m_\pi^2)_{\mbox{\rm\scriptsize ql}}
+
(m_\pi^2)_{\mbox{\rm\scriptsize ml}}
=
0+0=0~.
\end{equation}

\noindent The coefficients multiplying the three
(formally divergent) integrals in Eqs.~(\ref{QuarkLoops}) and (\ref{PionLoops})
are identically zero {\em before}
cutoffs keep these integrals finite.

    Extending this to SU(3) L$\sigma$M we get the GTRs:
$f_\pi\,g=\hat{m}$, $f_K\,g=\frac{1}{2}(m_s+\hat{m})$
where $\hat{m}=(m_u+m_d)/2$ along with the GTR ratio 1.22 from
data \cite{Eidelman:2004wy}

\begin{equation}
\frac{f_K}{f_\pi}
=
\frac{1}{2}\left(\frac{m_s}{\hat{m}}+1\right)\approx 1.22
\,\,\,\Rightarrow\,\,\, \frac{m_s}{\hat{m}}\approx 1.44~,
\label{fKoverfPi}
\end{equation}

\begin{eqnarray}
(m_K^2)_{\mbox{\rm\scriptsize ql}}
&=&
i\,4\,N_c\,g
\int\hat{d}^4p
\left(
-2g\frac{p^2-m_s\hat{m}}{(p^2-m_s^2)(p^2-\hat{m}^2)}
+\frac{g^\prime_{\mbox{\it\scriptsize ns}}}
      {m_{\sigma_{\mbox{\it\scriptsize ns}}}^2}
\frac{2\hat{m}}{p^2-\hat{m}^2}
+\sqrt{2}\frac{g^\prime_{\mbox{\it\scriptsize s}}}
      {m_{\sigma_{\mbox{\it\scriptsize s}}}^2}
\frac{m_s}{p^2-m_s^2}
\right)
\nonumber \\
&=&0\,\,\,\,\mbox{\rm in CL}
\end{eqnarray}

\noindent (see Ref.~\cite{Bramon:1997gg}, Eqs. (23) and (24),
or \cite{Delbourgo:1993dk}) leading via quark loops to

\begin{equation}
m_K^2
=
M^K_{\mbox{\rm\scriptsize VP}}
+
M^K_{\mbox{\scriptsize {\rm qktad},{\it ns}}}
+
M^K_{\mbox{\scriptsize {\rm qktad},{\it s}}}
=0
\end{equation}

\noindent in the CL (see Ref.~\cite{Delbourgo:1998kg}, Eq. (14a)).

\section{Pion and kaon masses away from CL}
\label{sec:PiAndKmasses}

The average nonstrange constituent quark mass is approximately

\begin{equation}
\hat{m}=\frac{1}{2}(m_u+m_d)
\approx\frac{M_N}{3}
\approx 313~\MeV~.
\label{mhat}
\end{equation}

\noindent
Equivalently, the low energy QCD
$1~\GeV$ scale $\langle - \bar{q}q\rangle\approx (245~\MeV)^3$
with the usual coupling $\alpha_s\approx 0.50$
suggests a dynamical mass \cite{Elias:1984zh,Elias:1984aw}

\begin{equation}
m_{\mbox{\rm\scriptsize dyn}}
=
\left( \frac{4\pi}{3} \alpha_s \langle - \bar{q}q\rangle \right)^{1/3}
\approx
313~\MeV~.
\end{equation}

\noindent Either the latter scale or Eq.~(\ref{mhat}) then predict
charge radii via the vector meson dominance (VMD) and
L$\sigma$M schemes

\begin{equation}
r_\pi^{\mbox{\rm\scriptsize VMD}}
=
\frac{\hj c\sqrt{6}}{m_\rho}
\approx 0.623~\mbox{\rm fm}~,
\end{equation}

\begin{equation}
r_\pi^{\mbox{\rm\scriptsize L$\sigma$M}}
=
\frac{\hj c}{\hat{m}}
\approx 0.63~\mbox{\rm fm}~,
\end{equation}

\noindent for $\hj c=197.3~\MeV\cdot\mbox{\rm fm}$
and with
$(\hat{m}+m_s)/2 = (337.5 + 486)/2 \approx 411.75~\MeV$
[see Eqs.~(\ref{m_s}) and (\ref{m_s_2})],

\begin{equation}
r_K^{\mbox{\rm\scriptsize VMD}}
=
\frac{\hj c\sqrt{6}}{m_K^\star}
\approx 0.54~\mbox{\rm fm}~,
\end{equation}

\begin{equation}
r_K^{\mbox{\rm\scriptsize L$\sigma$M}}
=
\frac{2 \hj c}{\hat{m}+m_s}
\approx 0.49~\mbox{\rm fm}~,
\end{equation}

\noindent near present data \cite{Eidelman:2004wy}

\begin{equation}
r_{\pi^+}
=
(0.672\pm 0.008)~\mbox{\rm fm}~,\,\,\,\,\,
r_{K^+}
=
(0.560\pm 0.031)~\mbox{\rm fm}~.
\end{equation}

\noindent Also the $u$ and $d$ constituent quark masses are
\cite{DeRujula:1975ge}

\begin{equation}
\hat{m}(\mbox{\rm mag. dipole moment})
=\frac{m_p}{2.792847}
\left[ 1 + \frac{14~\MeV}{9\hat{m}} \right]
\approx 337.5~\MeV~,
\label{mMagDipMom}
\end{equation}

\begin{equation}
m_u = 335.5~\MeV~,\,\,\,\,\,m_d=339.5~\MeV~,
\label{muANDmd}
\end{equation}

\noindent due to

\begin{equation}
m_d-m_u\simeq m_{K^0} - m_{K^+} = 3.97~\MeV~,
\end{equation}

\noindent away from the CL and isospin limit \cite{Delbourgo:1998qe}.
See Ref.~\cite{Scadron:2003yg} for a global $\bar{q}q$ picture of mesons.

   The quark-level L$\sigma$M predicts

\begin{equation}
N_c=3,\,\,\,\,m_\sigma=2m_q,\,\,\,\,\,
g=\frac{2\pi}{\sqrt{3}}
\approx 3.6276
\end{equation}

\noindent via either the L$\sigma$M \cite{Delbourgo:1993dk},
QCD in infrared limit \cite{Babukhadia:1997db}, $Z=0$
compositeness condition (Z=0 c.c.) \cite{SalamWeinberg,Scadron:1997nc}.
This implies via the GTR

\begin{equation}
\hat{m}=f_\pi\,g\approx 93~\MeV\cdot
\frac{2\pi}{\sqrt{3}}
\approx 337.4~\MeV~,
\end{equation}

\noindent very near (\ref{mMagDipMom}).
Given the above constituent quark masses away from the CL,
the chiral--breaking pion and kaon masses are found via
\cite{McKellar:1986dr} Eq.~(4.4).

  The difference between the constituent and dynamical quark mass
defines an effective current quark mass which vanishes in the CL
\cite{McKellar:1986dr}:

\begin{equation}
\delta \hat{m}
=
\hat{m}_{\mbox{\rm\scriptsize con}}
-
\frac{\hat{m}_{\mbox{\rm\scriptsize CL}}^3}
     {\hat{m}_{\mbox{\rm\scriptsize con}}^2}
=
337.5~\MeV-269.2~\MeV
=
68.3~\MeV~,
\label{DeltaHatM}
\end{equation}

\noindent where $\delta\hat{m}\to 0$ when
$\hat{m}_{\mbox{\rm\scriptsize con}}\to \hat{m}_{\mbox{\rm\scriptsize CL}}
\approx m_N/3
\approx 313~\MeV$.

\noindent Then because mesons are taken as $q\bar{q}$ states,

\begin{equation}
m_\pi = \delta\hat{m}+ \delta\hat{m} \approx 136.6~\MeV
\end{equation}

\noindent midway between $m_{\pi^+}=139.57~\MeV$ and
$m_{\pi^0}\approx 134.98~\MeV$
experimental masses \cite{Eidelman:2004wy}. Also
from Eq.~(\ref{fKoverfPi}) above,

\begin{equation}
m_s=1.44\, \hat{m}\approx 486.0~\MeV
\label{m_s}
\end{equation}

\noindent away from the CL or with $g=2\pi/\sqrt{3}$
and $f_K=1.22 f_\pi\approx 113.46~\MeV$,

\begin{equation}
m_s
=
2f_K\,g-\hat{m}
=
(823.2-337.5)~\MeV
\approx 485.7~\MeV
\label{m_s_2}
\end{equation}

\noindent close to (\ref{m_s}). However with
$m_s\approx 510~\MeV\approx m_\phi(1020)/2$ or via
magnetic dipole moments, one finds the average constituent quark masses
extending Eq.~(\ref{m_s}) to \cite{McKellar:1986dr}

\begin{equation}
m_s^{\mbox{\rm\scriptsize avg}}
=
(486+510)~\MeV/2
=
498~\MeV\,\,\,\,\Rightarrow
\end{equation}

\begin{equation}
\bar{m}
=
(498+337.5)~\MeV/2
=417.75~\MeV\,\,\,\,\Rightarrow
\end{equation}

\begin{eqnarray}
\delta^\prime \hat{m}
&=&
\hat{m}_{\mbox{\rm\scriptsize con}}
-\frac{\hat{m}_{\mbox{\rm\scriptsize CL}}^3}
      {\bar{m}^2}
\approx
(337.5-175.7)~\MeV
=
161.8~\MeV
\\
\delta^\prime m_s
&=&
m_{s,\mbox{\rm\scriptsize con}}^{\mbox{\rm\scriptsize avg}}
-\frac{\hat{m}_{\mbox{\rm\scriptsize CL}}^3}
      {\bar{m}^2}
\approx
(498-175.7)~\MeV
\approx
322.3~\MeV\,\,\,\,\Rightarrow
\end{eqnarray}

\begin{equation}
m_K
=
\delta^\prime \hat{m}
+
\delta^\prime m_s
=
(161.8+322.3)~\MeV
=
484.1~\MeV
\end{equation}

\noindent not far away from average \cite{Eidelman:2004wy} K mass
$m_{K^0}\approx 497.7~\MeV$ and $m_{K^+}\approx 493.7~\MeV$.

\section{$\eta_8$ mass}
\label{sec:Eta8}

In the CL $m_{\eta_8}=0$, consistent with
the {\em squared} Gell-Mann--Okubo (GMO) mass

\begin{equation}
m_{\eta_8}^2
=
\frac{4m_K^2-m_\pi^2}{3} \to 0\,\,\,\,\,\mbox{\rm in CL limit}~,
\label{meta8sqrdCL}
\end{equation}

\begin{equation}
\frac{f_8}{f_\pi}
=
\frac{3}{5-2\hat{m}/m_s}
\approx
0.831
\end{equation}

\noindent for $m_s/\hat{m}\approx 1.44$. ($f_8/f_\pi\to 1$ in the
U(3) symmetry limit.) 
Quark--level GTR for $\eta_8$ \cite{Eidelman:2004wy}
gives ($f_{\eta_8}/f_\pi) \approx 1.2$. Then

\begin{equation}
\sqrt{3} \frac{f_8}{f_\pi}
\left(\frac{f_{\eta_8}}{f_\pi}\right)
f_\pi\,g
=
\sqrt{3}
\cdot 0.831
\cdot 1.2
\cdot 92.42~\MeV
\cdot \frac{2\pi}{\sqrt{3}}
=
m_{\eta_8}
\approx
579.1~\MeV~,
\end{equation}

\noindent a dynamical estimate reasonably close to the GMO value
$m_{\eta_8}\approx 566.6~\MeV$ from Eq.~(\ref{meta8sqrdCL}).
Also $\eta_{\mbox{\it\scriptsize ns}}$ and
$\eta_{\mbox{\it\scriptsize s}}$
$\bar{q}q$ mixing masses are \cite{Klabucar:2001gr}
$m_{\eta_{\mbox{\it\tiny ns}}}\approx 758.1~\MeV$ and
$m_{\eta_{\mbox{\it\tiny s}}}\approx 801.2~\MeV$, respectively,
so that

\begin{equation}
m_\eta^2+m_{\eta^\prime}^2
=
m_{\eta_1}^2+m_{\eta_8}^2
=
m_{\eta_{\mbox{\it\tiny ns}}}^2 + m_{\eta_{\mbox{\it\tiny s}}}^2
\approx
1.217~\GeV^2~.
\end{equation}

  Reference \cite{Klabucar:2001gr} suggests using
$m_{\eta_8}=575.56~\MeV$ as in Eqs.~(\ref{thetaP}) and (\ref{meta8sqrd}) below

\begin{equation}
|\theta_P|
=
\arctan\sqrt{\frac{m_{\eta_8}^2-m_\eta^2}{m_{\eta^\prime}^2-m_{\eta_8}^2}}
\approx 13^\circ~,
\label{thetaP}
\end{equation}

\begin{equation}
m_{\eta_8}^2
=
\cos^2\theta_P\, m_\eta^2
+
\sin^2\theta_P\, m_{\eta^\prime}^2
\approx
(575.56~\MeV)^2~,
\label{meta8sqrd}
\end{equation}

\begin{equation}
\phi_P
=
\arctan\sqrt{\frac{m_{\eta_{\mbox{\it\tiny ns}}}^2-m_\eta^2}
                  {m_{\eta^\prime}^2-m_{\eta_{\mbox{\it\tiny ns}}}^2}}
\approx
41.84^\circ~,
\end{equation}

\begin{equation}
m_{\eta_{\mbox{\it\tiny ns}}}^2
=
\cos^2\phi_P\, m_\eta^2 + \sin^2\phi_P\, m_{\eta^\prime}^2
=
(758.1)^2~\MeV~,
\end{equation}

\begin{equation}
\theta_P
=
\phi_P
-
\arctan\sqrt{2}
\approx
41.84^\circ
-
54.74^\circ
=
-12.9^\circ~,
\end{equation}

\noindent close to $-13^\circ$ in Eq.~(\ref{thetaP})
and near the resulting GMO mass
$m_{\eta_8}\approx 566.6~\MeV$ in Eq.~(\ref{meta8sqrdCL}) away from the CL with
average

\begin{equation}
m_{\eta,\eta^\prime}^{\mbox{\rm\scriptsize avg}}
=
\frac{547.75+957.78}{2}~\MeV
=752.77~\MeV~,
\end{equation}

\noindent near $m_{\eta_{\mbox{\it\tiny ns}}}=758.1~\MeV$ above.
With hindsight, the above tightly bound
$q\bar{q}$ meson masses are near
data \cite{Eidelman:2004wy} in spite of their NG vanishing values.

\section{Chiral Goldberger--Treiman relations}
\label{sec:chiralGTRs}

  Given the massless NG pseudoscalars $m_\pi=m_K=m_{\eta_8}=0$ and their massive
version in Secs.~\ref{sec:PiAndKmasses} and \ref{sec:Eta8},
the massive chiral symmetry breaking poles combined with axial current
conservation then lead to the quark--level GTRs for pions and for kaons:

\begin{eqnarray}
f_\pi g &=& \hat{m} = \frac{1}{2}(m_u+m_d) \approx 337.5~\MeV~,
\label{fpig}
\\
f_K g &=& \frac{1}{2} (m_s + \hat{m}) \approx 411.8~\MeV~,
\label{fKg}
\end{eqnarray}

\noindent where we have invoked the constituent quark masses,
Eqs.~(\ref{mMagDipMom},\ref{muANDmd},\ref{m_s}). Also evaluating
the lhs of
Eqs.~(\ref{fpig}) and (\ref{fKg}) for $f_\pi\approx 93~\MeV$,
$f_K\approx 1.22 f_\pi\approx 113.5~\MeV$ and the meson--quark coupling
$g\approx 2\pi/\sqrt{3}\approx 3.6276$ for $N_c=3$
\cite{Delbourgo:1993dk,Elias:1984zh,Elias:1984aw},
the lhs of Eqs.~(\ref{fpig},\ref{fKg}) becomes $337.4~\MeV$,
$411.6~\MeV$, in very
close agreement with the rhs of Eqs.~(\ref{fpig},\ref{fKg}), respectively.

   For finite UV cutoff $\Lambda$, the pion coupled to the axial current via
the quark loop with $g=2\pi/\sqrt{N_c}$ leads to

\begin{equation}
\int
\frac{d^4p}{(p^2-m_q^2)^2}
=
i\pi^2~,
\end{equation}

\noindent or equivalently to the log--divergent
gap equation \cite{Delbourgo:1993dk,Scadron:2002mm}

\begin{equation}
-i4N_c g^2
\int \frac{\hat{d}^4p}{(p^2-m_q^2)^2} = 1~.
\label{gapeq}
\end{equation}

\noindent Explicitly accounting for $\Lambda$, the lhs of Eq.~(\ref{gapeq})
can be written as \cite{Delbourgo:1993dk}

\begin{equation}
\ln\left(1+\frac{\Lambda^2}{m_q^2}\right)
-\frac{1}{1+\frac{m_q^2}{\Lambda^2}}=1~,
\end{equation}

\noindent with the numerical solution

\begin{equation}
\frac{\Lambda^2}{m_q^2}\approx (2.3)^2~,
\label{LambdaScale}
\end{equation}

\noindent or

\begin{equation}
\Lambda^{\mbox{\rm\scriptsize CL}}
\approx
2.3\,\hat{m}^{\mbox{\rm\scriptsize CL}}_q
\approx 750~\MeV~,
\label{LambdaScaleCL}
\end{equation}

\noindent for CL quark mass $325.7~\MeV$ and
$\Lambda\approx 2.3\, \hat{m}_q \approx 776~\MeV$ for chiral--broken mass
$337.5~\MeV$.

   It is significant that the above quark mass scales from
Eq.~(\ref{LambdaScaleCL}) correspond to the
Z=0 c.c. \cite{SalamWeinberg,Scadron:1997nc} with
$\Lambda^{\mbox{\rm\scriptsize CL}}<\Lambda\approx 776~\MeV$
near the $\rho(775)$ and $\omega(782)$ slightly bound $\bar{q}q$ masses,
but slightly heavier then $\Lambda^{\mbox{\rm\scriptsize CL}}\approx 750~\MeV$.
In a similar fashion, the $\Lambda^\prime$ cutoff for the vector $K^\star$
$\bar{q}q$ mass is (for $m_s\approx 469~\MeV$ and
$\hat{m}\approx 325.7~\MeV$ see Eq.~(\ref{mhatCL}) below)
in the CL:

\begin{equation}
\Lambda^\prime \approx 2.3 \sqrt{m_s\,\hat{m}}\approx 899~\MeV~,
\end{equation}

\noindent reasonably near the observed $K^\star(894)$ mass.
Lastly, the Z=0 c.c. also requires the meson--quark coupling to be
\cite{Scadron:1997nc} $g\approx 2\pi/\sqrt{3}$, analogous to the infrared
QCD limit \cite{Elias:1984zh,Elias:1984aw} and also found
via the quark--level L$\sigma$M \cite{Delbourgo:1993dk}.


\section{Extension to $\bar{q}q$ scalar and vector masses}
\label{sec:ScalarAndVector}

To complete the ground state $\bar{q}q$ meson scheme,
we summarize and update the results of Ref.~\cite{Scadron:2003yg},
first obtaining the SU(3) ground state octet vector meson
$\bar{q}q$ masses from the bare plus symmetry--breaking terms
as $m_V=\sqrt{2/3}\,m_V^0-d_{i8i}\,\delta m_V$:

\begin{eqnarray}
m_{\rho,\omega}
=
\sqrt{\frac{2}{3}} m_V^0
-
\frac{1}{\sqrt{3}}\delta m_V
\approx
779~\MeV~,
\label{mRhoOmega}
\\
m_{K^\star}
=
\sqrt{\frac{2}{3}} m_V^0
+
\frac{1}{2\sqrt{3}}\delta m_V
\approx
894~\MeV~,
\\
m_\phi
=
\sqrt{\frac{2}{3}} m_V^0
+
\frac{2}{\sqrt{3}}\delta m_V
\approx
1019~\MeV~,
\label{mPhi}
\end{eqnarray}

\noindent leading to the average scales

\begin{equation}
m_V^0 \approx 961~\MeV~,\,\,\,\,\,
\delta m_V\approx 139~\MeV~,\,\,\,\,\,
\frac{\delta m_V}{m_V^0} \approx 14\%~.
\label{AverageScales}
\end{equation}

\noindent Also for scalar meson masses, the model--independent nonstrange
scalar sigma mass is \cite{Surovtsev:2002kr}

\begin{equation}
m_{\sigma_{\mbox{\it\scriptsize ns}}}
\approx 665~\MeV~.
\label{mSigmaNS}
\end{equation}

\noindent This can be verified by first working in the CL with
NJL--L$\sigma$M mass

\begin{equation}
m_\sigma^{\mbox{\rm\scriptsize CL}}
=
2\hat{m}^{\mbox{\rm\scriptsize CL}}
\approx
651.4~\MeV~,
\label{NJL-LsM-mass}
\end{equation}

\noindent due to the GTR. In the CL

\begin{equation}
\hat{m}^{\mbox{\rm\scriptsize CL}}
=
f_\pi^{\mbox{\rm\scriptsize CL}}\, g
=
(89.775~\MeV)
\frac{2\pi}{\sqrt{3}}
\approx
325.7~\MeV~,
\label{mhatCL}
\end{equation}

\noindent leading to (\ref{NJL-LsM-mass}). Also
$f_\pi^{\mbox{\rm\scriptsize CL}}$ above follows from the once--subtracted
dispersion relation \cite{Nagy:2004tp}

\begin{equation}
\frac{f_\pi}{f_\pi^{\mbox{\rm\scriptsize CL}}}-1
=
\frac{m_\pi^2}{8\pi^2 f_\pi^2}
\left( 1 + \frac{m_\pi^2}{10\,\hat{m}^2} \right)
\approx
2.946\%~,
\end{equation}

\noindent giving the observed pion decay constant \cite{Eidelman:2004wy}
-- extended in the CL as

\begin{equation}
f_\pi\approx 92.42~\MeV~,\,\,\,\,\,\,\,\,\,
f_\pi^{\mbox{\rm\scriptsize CL}}=
\frac{f_\pi}{1.02946}
\approx 89.775~\MeV~.
\end{equation}

\noindent Then the nonstrange scalar mass satisfies

\begin{equation}
m_{\sigma_{\mbox{\it\scriptsize ns}}}^2
-
m_\pi^2
=
(m_\sigma^{\mbox{\rm\scriptsize CL}})^2\,\,\,\,\,
\mbox{\rm or}\,\,\,\,\,
m_{\sigma_{\mbox{\it\scriptsize ns}}}
\approx 665.82~\MeV~,
\end{equation}

\noindent compatible with (\ref{mSigmaNS}). Also the scalar kappa mass satisfies

\begin{equation}
m_\kappa
=
2\sqrt{\hat{m}m_s}
\approx 810~\MeV~
\end{equation}

\noindent for $\hat{m}\approx 337.5~\MeV$ and from (\ref{fKoverfPi}),
$m_s\approx 1.44\,\hat{m}\approx 486~\MeV$, compatible with E791 data
\cite{Aitala:2002kr} $m_\kappa=797\pm 19~\MeV$.
Finally, the pure strange scalar mass satisfies

\begin{equation}
m_{\sigma_{\mbox{\rm\scriptsize S}}}
\simeq
2 m_s
\approx
972~\MeV~,
\end{equation}

\noindent near the almost pure $\bar{s}s$ vector mass $m_\phi\approx 1019~\MeV$.

   Then the scalar analog of the SU(3) vector masses
(\ref{mRhoOmega})--(\ref{mPhi}) are

\begin{eqnarray}
m_{\sigma_{\mbox{\it\scriptsize ns}}}
=
\sqrt{\frac{2}{3}} m_s^0
-
\frac{1}{\sqrt{3}}\,\delta m_s
\approx 665~\MeV~,
\label{mSigmaNonStrange}
\\
m_\kappa
=
\sqrt{\frac{2}{3}} m_s^0
+
\frac{1}{2\sqrt{3}}\,\delta m_s
\approx 810~\MeV~,
\\
m_{\sigma_{\mbox{\it\scriptsize s}}}
=
\sqrt{\frac{2}{3}} m_s^0
+
\frac{2}{\sqrt{3}}\,\delta m_s
\approx 972~\MeV~,
\label{mSigmaStrange}
\end{eqnarray}

\noindent giving the average SU(3) scalar $\bar{q}q$ masses,

\begin{equation}
m_s^0\approx 933~\MeV~,\,\,\,\,\,
\delta m_s\approx 177~\MeV~,\,\,\,\,\,
\frac{\delta m_s}{m_s^0}
\approx
19\%~,
\end{equation}

\noindent reasonably near the average SU(3) vector $\bar{q}q$ masses in
(\ref{AverageScales}).

   Also given the closeness of these scalar masses in
(\ref{mSigmaNonStrange})--(\ref{mSigmaStrange}) to the approximate quark mass
sums in Tab.~\ref{tab:QuarkScalarMasse}, all scalar meson masses have
essentially ``touching quarks''. However, the vector masses in
(\ref{mRhoOmega})--(\ref{mPhi}) are ``loosely bound quarks''
as they are 115 to 50 MeV heavier then the above ``touching quark''
scalar meson masses. In Ref.~\cite{McKellar:1986dr}
a nonrelativistic quark model
$\vect{L}\cdot\vect{S}$ coupling roughly accounts for the difference
between vector and scalar meson masses.

\begin{table}
\begin{tabular}{|c|c|c|}
\hline
 & data Ref.~\cite{Eidelman:2004wy} & quark sum \\
\hline
$\sigma_{\mbox{\it\scriptsize ns}}$ &
665 &
$2\hat{m}\approx 675$ \\
$\kappa$ &
$(797\pm 19)$ &
$2\sqrt{\hat{m}m_s}\approx 810$ \\
$\sigma_s$ &
$(980\pm 10)$ &
$2m_s\approx 972$ \\
\hline
\end{tabular}
\caption{Touching--quark scalar meson masses (in $\MeV$) for
$\hat{m}\approx 337.5~\MeV$ and $m_s\approx 486~\MeV$.
}
\label{tab:QuarkScalarMasse}
\end{table}


\section{Conclusion}
\label{sec:conclusion}

In this paper we have primarily focused on the ground state $\bar{q}q$
pseudoscalar mesons, which we model via the quark--level L$\sigma$M,
briefly described in Sec.~\ref{sec:QL_LsM}.

   The vanishing chiral NG pion and kaon masses are discussed in
Sec.~\ref{sec:NGtheorem}. Their non--vanishing mass values away from
the CL are discussed in Sec.~\ref{sec:PiAndKmasses}, characterized
by the tightly--bound $\bar{q}q$ charge radii $\hj c/\hat{m}$,
$2\hj c/(\hat{m}+m_s)$ for charged pions and kaons, respectively.
The latter are also close to the VMD values. In Sec.~\ref{sec:Eta8}
we have extended the vanishing NG $\eta_8$ mass to its non--vanishing
GMO and its tightly bound $\bar{q}q$ meson--mixing value.
Sec.~\ref{sec:chiralGTRs} deals with chiral--symmetric pion and kaon GTRs.
In Sec.~\ref{sec:ScalarAndVector} we extended the L$\sigma$M scheme
to $\bar{q}q$ vector and scalar masses.
Note that the loosely bound SU(3) masses $m_S^0\approx 933~\MeV$ and
$m_V^0\approx 961~\MeV$ are close to the tightly bound $m_P$
mass characterized by the observed \cite{Eidelman:2004wy} $\eta^\prime$ mass
$m_{\eta^\prime}=(957.78\pm 0.14)~\MeV$.

   The above $\bar{q}q$ scheme appears somewhat counter
to the PDG p.848 ``non--$\bar{q}q$ candidates'' \cite{nonqbarq}.
However, in Ref.~\cite{Scadron:2003yg} we remind the reader of the standard
ground state SU(3) $qqq$ octet and decuplet baryon states with
\cite{Scadron:2003yg} $m_O^0\sim 1150~\MeV$, $m_D^0\sim 1230~\MeV$
being about $250~\MeV$ greater then the above $\bar{q}q$ ground states
$m_V^0\sim m_S^0\sim m_{\eta^\prime} \sim 950~\MeV$.
Taking $\phi\sim \bar{s}s$ or $m_s\sim 500~\MeV$
and $J/\psi(3100)\sim \bar{c}c$
or $m_c\sim 1550~\MeV$, one can reasonably model the (higher mass) ground state
$qqq$ baryon states vs. data \cite{Eidelman:2004wy} as in Tab.~\ref{tab:qqq}.

   With hindsight, the recent paper \cite{kekez05mod} complements
the present L$\sigma$M $q\bar{q}$ picture quite well.
In particular, Ref.~\cite{kekez05mod} shows in
detail why the quark triangle L$\sigma$M predictions (involving no arbitrary
parameters for at least 15 decays) match data \cite{Eidelman:2004wy} to within 5\%.
This is for $V\to PV$ or $P\to VV$ strong or electromagnetic
decays. Also, the dynamic Schwinger--Dyson approach describes $\pi$, $K$, and
$\eta$ decays in conformity with empirical constraints.

\begin{table}
\begin{tabular}{|c|c|c|c|}
\hline
\multicolumn{2}{|c|}{\begin{tabular}{c} constituent quark\vspace*{-1mm} \\ predictions\end{tabular}} & \multicolumn{2}{c|}{data from \cite{Eidelman:2004wy} } \\
\hline
$qqq$ & $m[\MeV]$ & & $m[\MeV]$ \\
\hline
$sdd$ & $1180$ & $\Sigma^-$    & $1197.45\pm 0.30$ \\
$ssd$ & $1340$ & $\Xi^-$       & $1321.31\pm 0.13$ \\
$sss$ & $1500$ & $\Omega^-$    & $1672.45\pm 0.29$ \\
$cud$ & $2225$ & $\Lambda_c^+$ & $2284.9\pm 0.6$   \\
$csu$ & $2390$ & $\Xi_c^+$     & $2466.3\pm 1.4$   \\
$css$ & $2550$ & $\Omega_c^0$  & $2697.5\pm 2.6$   \\
$ccd$ & $3440$ & $\Xi_{cc}^+$  & $3519\pm 1$       \\
\hline
\end{tabular}
\caption{Loosely bound $qqq$ baryons (in MeV) for $m_u\approx 335.5~\MeV$
and $m_d\approx 339.5~\MeV$, $m_s\approx 500~\MeV$, and $m_c\approx 1550~\MeV$.}
\label{tab:qqq}
\end{table}


\end{document}